\documentclass{blois}

\usepackage{multirow}
\usepackage{setspace}
\usepackage{hhline}

\def\Journal#1#2#3#4{{#1} {\bf #2}, #3 (#4)}

\def\PLB{{\em Phys. Lett.}  B}
\def\PRL{\em Phys. Rev. Lett.}
\def\PRD{{\em Phys. Rev.} D}

\def\mco{\multicolumn}

\def\be{\begin{equation}}
\def\ee{\end{equation}}
\def\bea{\begin{eqnarray}}
\def\eea{\end{eqnarray}}

\newcommand{\Photo}{\includegraphics[height=35mm]{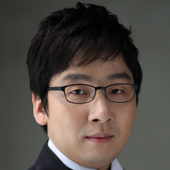}}

\begin{document}
\vspace*{4cm}
\title{PRODUCTION OF SINGLE TOP QUARK - RESULTS FROM THE TEVATRON AND THE LHC}

\author{ CHANG-SEONG MOON }

\address{AstroParticule et Cosmologie, Universit\'e Paris Diderot, CNRS/IN2P3,\\
10, rue Alice Domon et L\'eonie Duquet 75205 Paris, France\\
and INFN Sezione di Pisa, Largo B. Pontecorvo, 3, 56127 Pisa, Italy}

\maketitle\abstracts{
We present the most recent measurements of single top quark production cross section by the CDF and D0 experiments 
at the Fermilab Tevatron Collider and the ATLAS and CMS experiments at the Large Hadron Collider (LHC).
The data were collected at the Tevatron corresponding to an integrated luminosity of up to 9.7 fb$^{-1}$ of 
proton-antiproton ($p\bar p$) collisions at a centre-of-mass energy of 1.96 TeV and at the LHC corresponding to an integrated 
luminosity of up to 4.9 fb$^{-1}$ of proton-proton ($pp$) collisions at a centre-of-mass energy of 7 TeV in 2011 and up to 
20.3 fb$^{-1}$ at a centre-of-mass energy of 8 TeV in 2012.
The measurements of single top quark production in $s$-channel, $t$-channel and associated production of a top quark 
and a $W$-boson ($tW$ production) are presented separately and lower limits on the CKM matrix element $|V_{tb}|$ 
from the single top quark cross section are set.
}

\section{Introduction}

\begin{figure}
\begin{minipage}{0.33\linewidth}
\centerline{\includegraphics[width=0.7\linewidth]{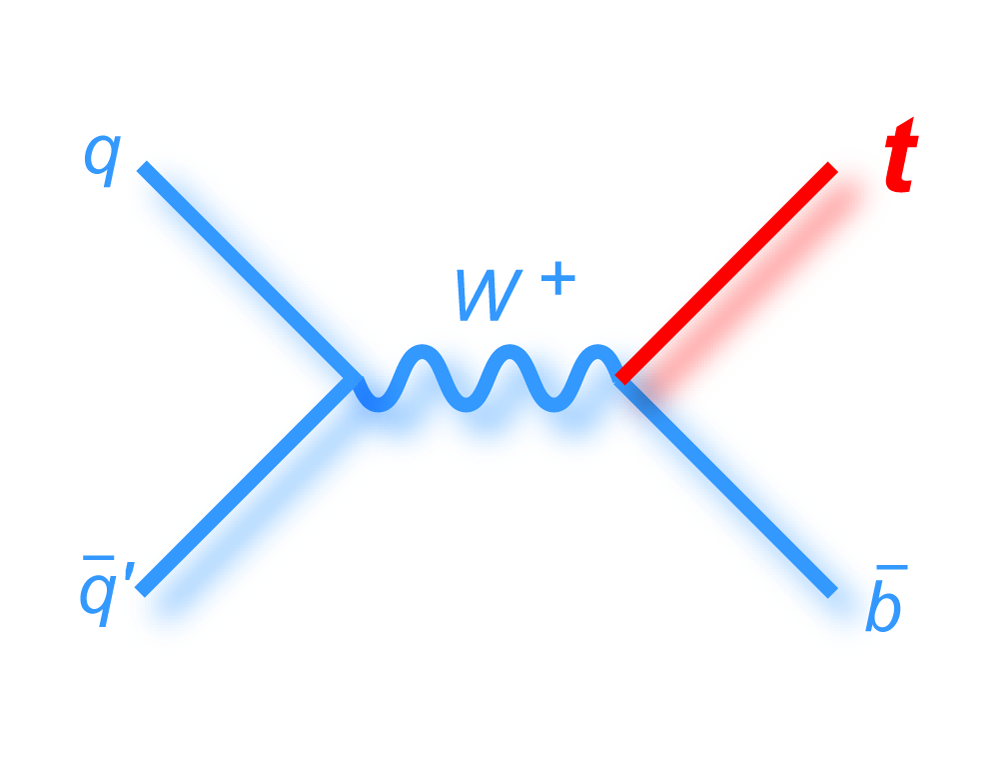}}
\end{minipage}
\hfill
\begin{minipage}{0.32\linewidth}
\centerline{\includegraphics[width=0.7\linewidth]{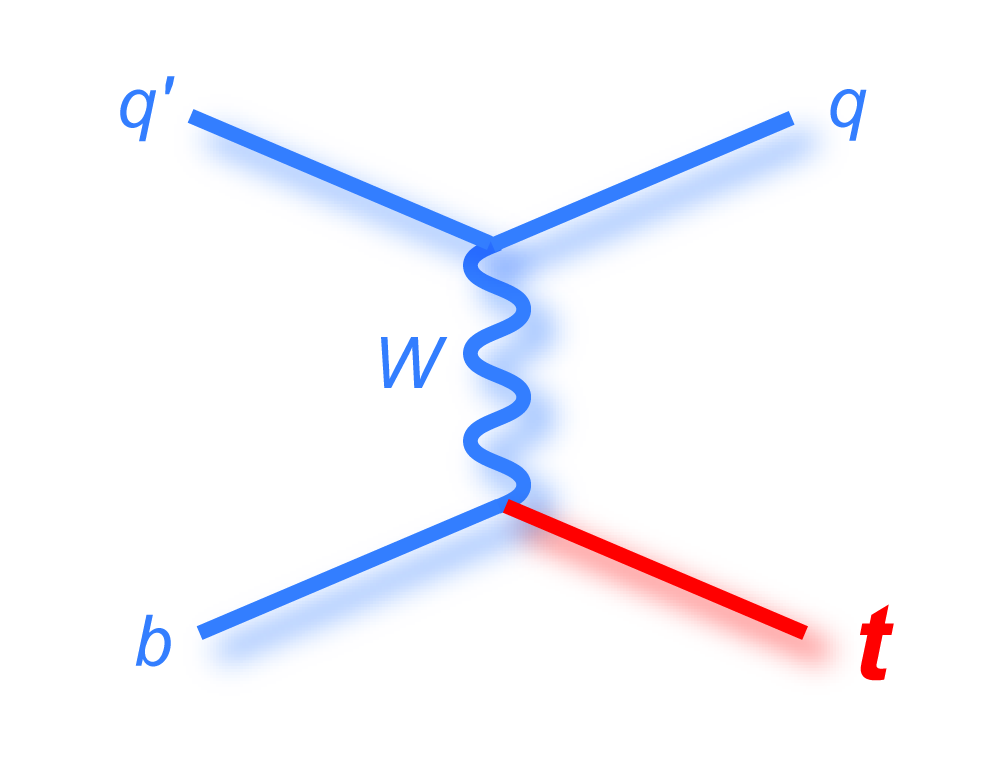}}
\centerline{\includegraphics[width=0.7\linewidth]{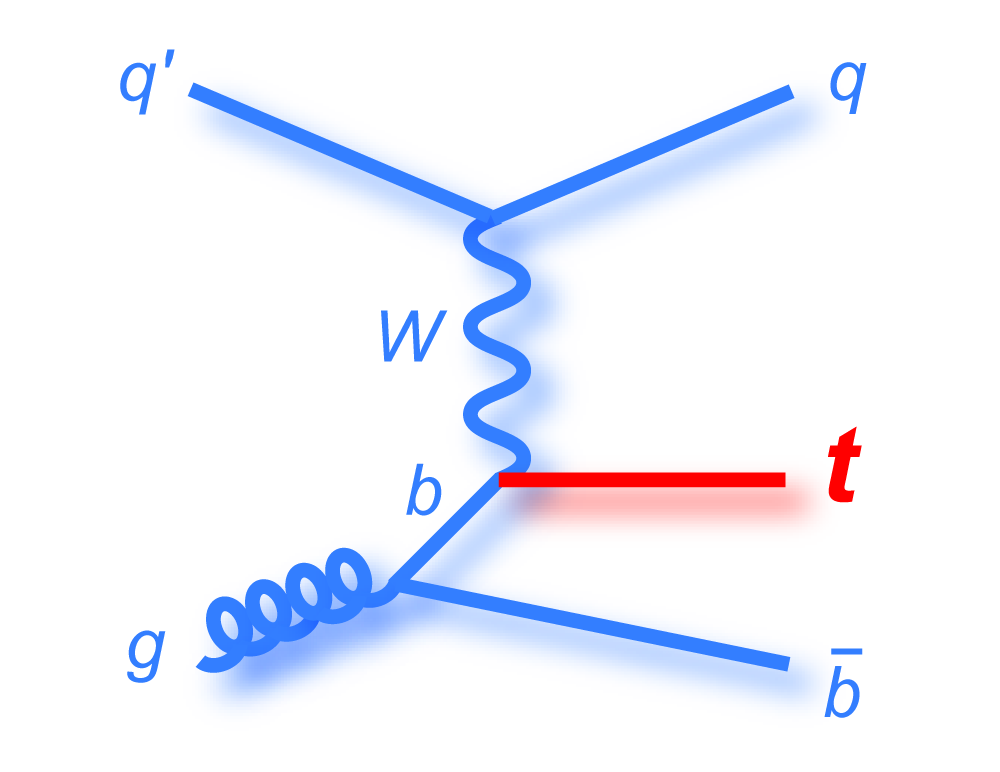}}
\end{minipage}
\hfill
\begin{minipage}{0.32\linewidth}
\centerline{\includegraphics[width=0.7\linewidth]{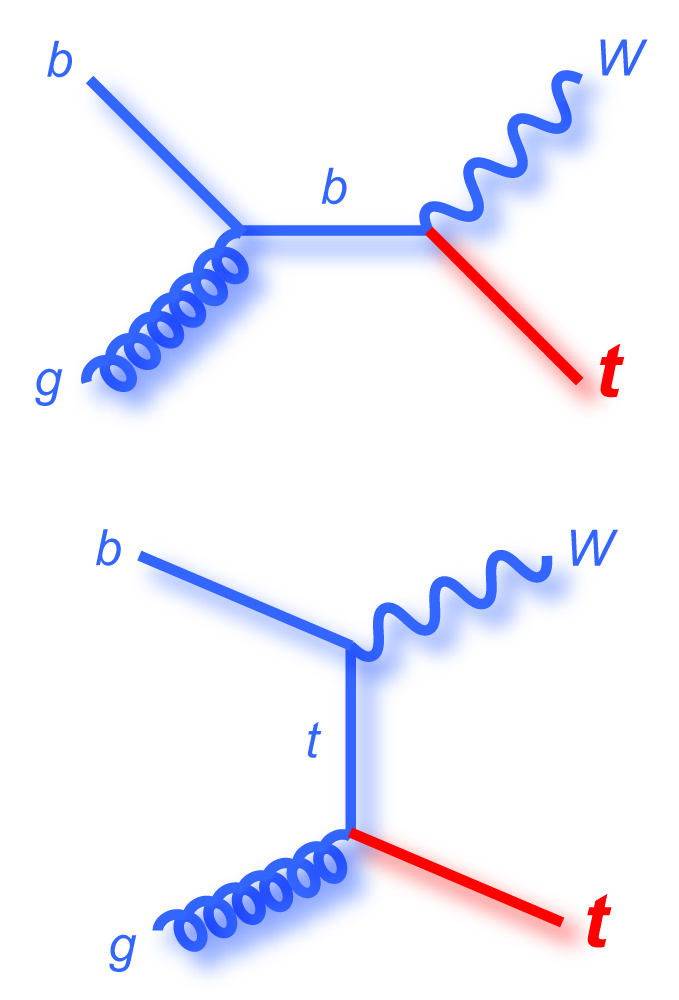}}
\end{minipage}
\caption[]{
Feynman diagrams of single top quark production for $s$-channel (left), $t$-channel (center) and $tW$ production 
(right) respectively}
\label{fig:channel}
\end{figure}

Top quarks are predominantly produced in pairs via the strong interaction in $pp$ or $p\bar{p}$ collisions 
but they can be also are produced singly via the electroweak interaction. 
The Standard Model (SM) predicts three single top quark processes which are $t$-channel, $s$-channel and associated 
production of a top quark and a $W$-boson which are shown in Figure~\ref{fig:channel}.
Main challenge to observe the single top process, is to overcome large background for extraction of the single 
top signal.
The single top quark production was first observed by the CDF~\cite{CDF_observe} and D0~\cite{D0_observe} 
experiments at the Tevatron in 2009. 
Recently many new results on the single top production have been reported from the Tevatron and the LHC.
We will summarize and discuss the most important results on the ATLAS, CDF, CMS and D0 experiments in this paper.

\section{Physics Motivation}

Single top production is an important background for the SM Higgs boson production but also physics process to 
test the SM prediction via directly measuring the CKM matrix element $|V_{tb}|$. 
Any excess of $|V_{tb}|$ over the SM prediction indicates the presence of new physics process beyond the SM. 
Also the $t$-channel single top production cross section provides a test of the $b$ parton distribution 
function of the proton.
In addition, single top production cross section is sensitive to new physics such as a fourth generation of quarks, 
flavor-changing $Z$-$t$-$c$ coupling (e.g. production of $p\bar{p}\rightarrow t\bar{c}$),
additional charged gauge boson ($W^\prime$), charged Higgs boson production or other new phenomena~\cite{cdfd0}.

\section{Single Top Production at the Tevatron}

The $s$- and $t$-channel processes of single top production are dominant at the Tevatron 
but almost negligible $tW$ production cross section in $p\bar p$ collisions. 
The next-to-next-to-leading-order (NNLO) cross section for $s$- and $t$-channel processes were calculated respectively to be  
$\sigma_s$ = 1.05 $\pm$ 0.06 pb~\cite{tevatron_sigma_s} and $\sigma_t$ = 2.10 $\pm$ 0.13 pb~\cite{tevatron_sigma_t}
assuming a top quark mass of 172.5 GeV. 
Since the predicted $s$-channel production cross section is smaller than that of $t$-channel, the observation of $s$-channel
process is more difficult. However, There is an advantage in $s$-channel mode at the Tevatron comparing to the LHC
since valence quarks ($q\bar q^\prime$ from $p\bar p$) generally initiate $s$-channel single top quark production,
leading to a larger signal-to-background ratio at the Tevatron than at the LHC~\cite{cdfd0}.

\begin{figure}[htbp]
\hspace{20mm} (a) \hspace{34mm} (b) \hspace{32mm} (c) \hspace{27mm} (d)
\begin{center}
\includegraphics[width=41mm,height=39mm]{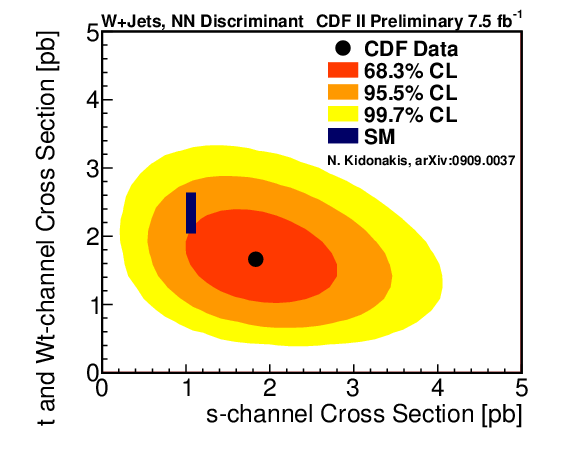}
\includegraphics[width=39mm,height=39mm]{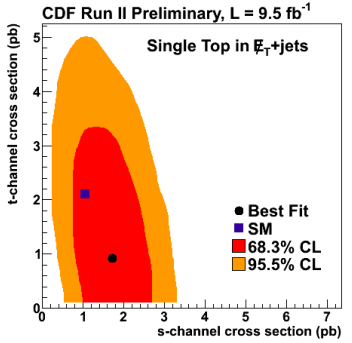}
\includegraphics[width=37mm,height=37mm]{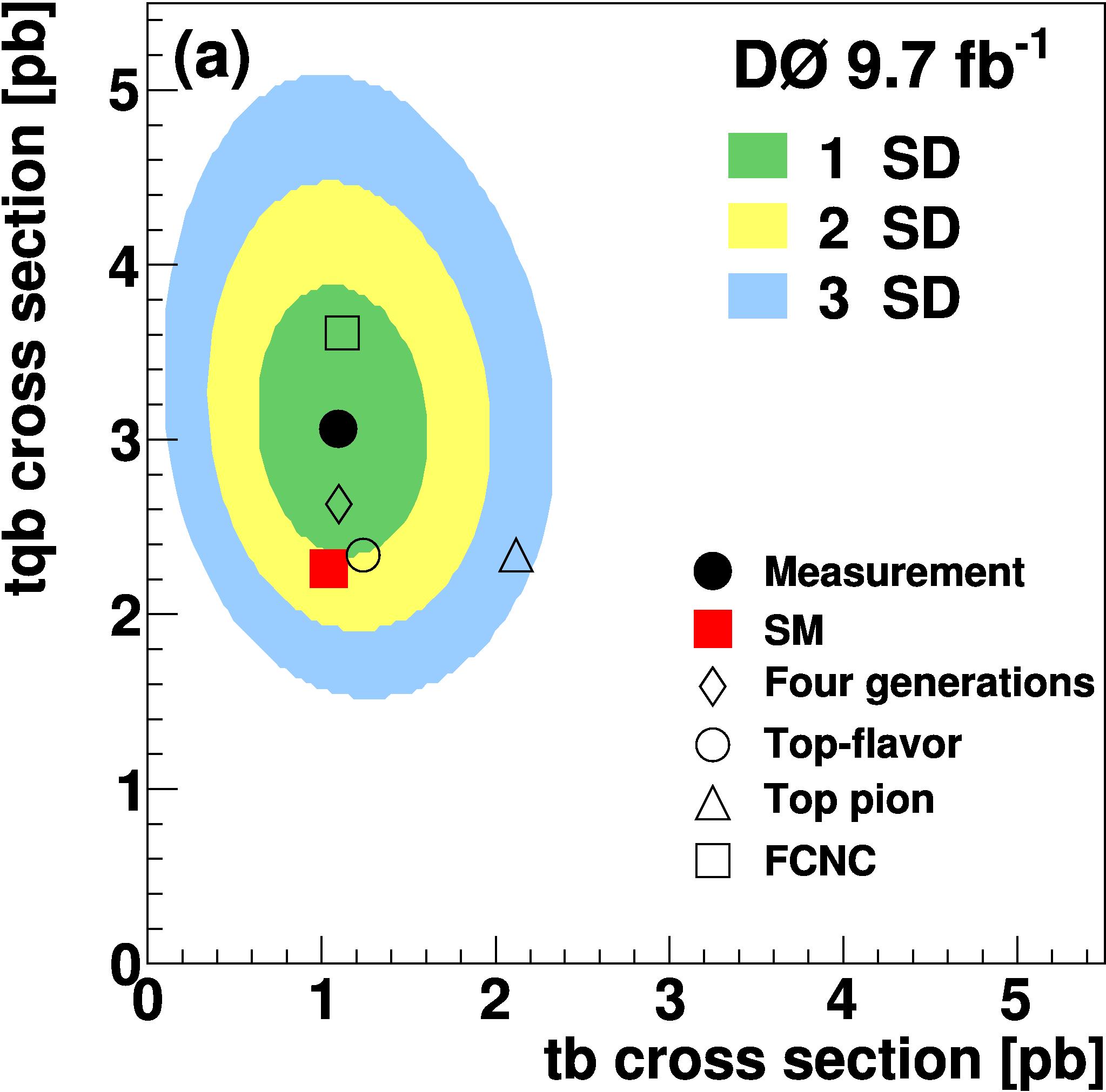}
\includegraphics[width=37mm,height=37mm]{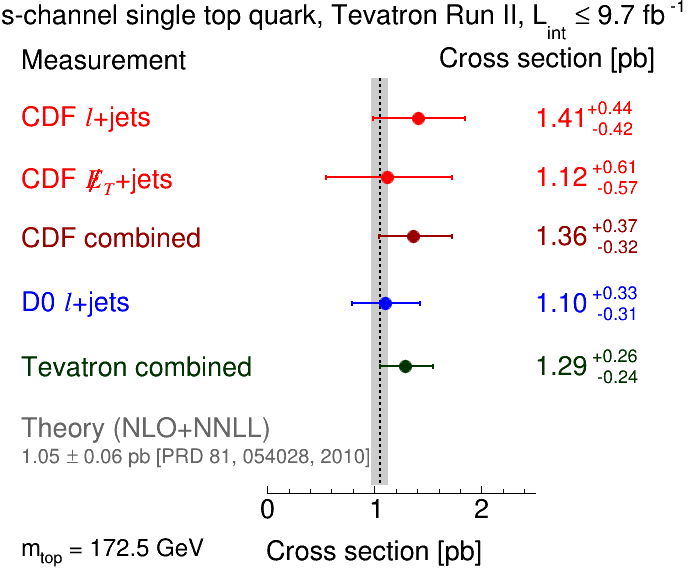}
\end{center}
\caption[]{
Results of the two-dimensional fit for $s$- and $t$-channel production cross section for 
(a)  $l$+jets events at CDF,
(b)  missing $E_T$+jets events at CDF,
(c) $l$+jets events at D0,
and (d) summary of measured $s$-channel production cross sections from each of the individual analyses by CDF and D0 experiments.}
\label{fig:tevatron}
\end{figure}

Measurements of Single top production in the combined $s$- and $t$-channels was performed by CDF and D0 experiments.
Two independent measurements were reported from CDF experiments. 
The first measured single top production cross section is shown in Figure~\ref{fig:tevatron}(a) to be 
$3.04^{+0.57}_{-0.53}$ pb ($\sigma_s = 1.81^{+0.63}_{-0.58}$ pb, 
$\sigma_t = 1.49^{+0.47}_{-0.42}$ pb) for $m_t$ = 172.5 GeV and $|V_{tb}|$ = 0.96 $\pm$ 0.09(stat.+syst.) $\pm$ 0.05(theory),
while lower limit of $|V_{tb}|> $ 0.78 is set at 95\% confidence level (CL)
using events with one charged lepton, large missing transverse energy, and jets~\cite{cdf_7.5fb}. 
And the second measurement of the cross section is $3.53^{+1.25}_{-1.16}$ pb and lower limit of $|V_{tb}| > $ 0.63 at 95\% CL
using missing transverse energy plus jets topology as shown in Figure~\ref{fig:tevatron}(b).

The D0 collaboration presented the observation of single top production in $t$-channel and first evidence for 
$s$-channel process~\cite{D0_s}.
The obtained production cross section in the combined channel is $4.11^{+0.59}_{-0.55}$ pb ($m_t$ = 172.5 GeV) 
in agreement with the SM prediction of $3.34^{+0.53}_{-0.49}$ pb and set lower limit of  $|V_{tb}| > $ 0.92 at 95\% CL.
And Figure~\ref{fig:tevatron}(c) shows the simultaneous two-dimensional fit measurements to be 
$\sigma_s$ = $1.10^{+0.33}_{-0.31}$ pb corresponding to a significance of 3.7 standard deviations ($\sigma$) (3.7$\sigma$ expected),
$\sigma_t$ = $3.07^{+0.53}_{-0.49}$ pb corresponding to a significance of 7.7$\sigma$ (6.0$\sigma$ expected) 

The CDF collaboration confirmed the evidence for $s$-channel single top production using two independent final states.
The $s$-channel single top production cross sections were measured respectively 
$\sigma_s$ = $1.41^{+0.44}_{-0.42}$ pb with a significance of 3.8$\sigma$ (2.9$\sigma$ expected)~\cite{cdf_s_lvbb} 
using lepton plus jets events and 
$\sigma_s$ = $1.12^{+0.61}_{-0.57}$ pb with a significance of 1.9$\sigma$ (1.8$\sigma$ expected)~\cite{cdf_s_vbb} 
using missing transverse energry plus jets events, 
both for $m_t$ = 172.5 GeV.

Recently the first observation of $s$-channel single top production with the combination of the CDF and D0 measurements 
of the production cross section was reported.
The measured production cross section is $\sigma_s$ = $1.29^{+0.26}_{-0.142}$ pb 
and a significance of standard deviation is 6.3$\sigma$ for the presence of an $s$-channel contribution to the production 
of single top quarks~\cite{cdfd0}.
The summary of $s$-channel production cross section measurements are shown in Figure~\ref{fig:tevatron}(d).

\section{Single Top Production at the LHC}

The $t$-channel and $tW$ production processes are dominant at the LHC but $s$-channel not reachable yet.
The NNLO cross section for three single top processes assuming $m_t$ = 172.5 GeV are calculated 
as shown in Table~\ref{tab:lhc}~\cite{tevatron_sigma_s,tevatron_sigma_t,lhc_sigma}.

\begin{table}[htbp]
\setstretch{1.20}
\centering
\caption[]{NNLO cross section calculation for three single top production processes at the LHC}
\label{tab:lhc}
\vspace{0.4cm}
\begin{tabular}{|c|c|c|c|c|} \hhline{--~--}
LHC (7TeV)      & $\sigma_t$ + $\sigma_{\bar t}$ (pb)  & & LHC (8TeV)      & $\sigma_t$ + $\sigma_{\bar t}$ (pb)  \\\hhline{--~--}
$t$-channel     & $65.9^{+2.1}_{-0.7}$$^{+1.5}_{-1.7}$ & & $t$-channel     & $87.2^{+2.8}_{-1.0}$$^{+2.0}_{-2.2}$ \\\hhline{--~--}
$s$-channel     & $4.56 \pm 0.07^{+0.18}_{-0.17}$      & & $s$-channel     & $5.55 \pm 0.08 \pm 0.21$             \\\hhline{--~--}
$tW$ production & $15.6 \pm 0.4 \pm 1.1$               & & $tW$ production & $22.2 \pm 0.6 \pm 1.4$               \\\hhline{--~--}

\end{tabular}
\end{table}

New measurements of $t$-channel single top production are obtained using full statistics at 8 TeV
from ATLAS and CMS experiments independently.
A value of $\sigma_t$ = 82.6 $\pm$ 1.7(stat.) $\pm$ 11.4(syst.) $\pm$ 3.1(PDF) $\pm$ 2.3(lumi.) pb for $m_t$ = 172.5 GeV is measured
and the coupling strength at the $W$-$t$-$b$ vertex is determined to be $|V_{tb}|$ = 0.97$^{+0.09}_{-0.10}$,
lower limit of $|V_{tb}| > $ 0.78 at 95\% CL is set by ATLAS experiment.
The obtained production cross section from CMS experiment is $\sigma_t$ = 83.6 $\pm$ 2.3(stat.) $\pm$ 7.4(syst.) pb 
for $m_t$ = 173 GeV and $|V_{tb}|$ is = 0.98 $\pm$ 0.05(experiment) $\pm$ 0.02(theory), 
lower limit of $|V_{tb}| > $ 0.92 at 95\% CL~\cite{cms_t}.

Evidence for $tW$ production process was reported by ATLAS experiment with full data set at 8 TeV 
and the measured production cross section is 
$\sigma_{tW}$ = 27.2 $\pm$ 2.8(stat.) $\pm$ 5.4(syst.) pb with a significance of 4.2 $\sigma$ (4.0$\sigma$ expected).
The $W$-$t$-$b$ vertex is determined to be $|V_{tb}|$ = 1.10 $\pm$ 0.12(experiment) $\pm$ 0.03(theory),
lower limit of $|V_{tb}| > $ 0.72 at 95\% CL.
CMS collaboration presented the first observation of the $tW$ single top production process, 
assuming $m_t$ = 172.5 GeV the measured production cross section is $\sigma_{tW}$ = 23.4 $\pm$ 5.4 pb in agreement with 
the SM of 22.2 $\pm$ 0.6 $\pm$ 1.4 pb corresponding to a significance of 6.1 $\sigma$ (5.4$\sigma$ expected),
The CKM matrix element is obtained to be $|V_{tb}|$ = 1.03 $\pm$ 0.12(experiment) $\pm$ 0.04(theory) and 
to be set lower limit of $|V_{tb}| > $ 95\% CL~\cite{cms_tw}.

\begin{figure}[htbp]
\hspace{27mm} (a) \hspace{44mm} (b) \hspace{42mm} (c)
\begin{center}
\includegraphics[width=52mm,height=45mm]{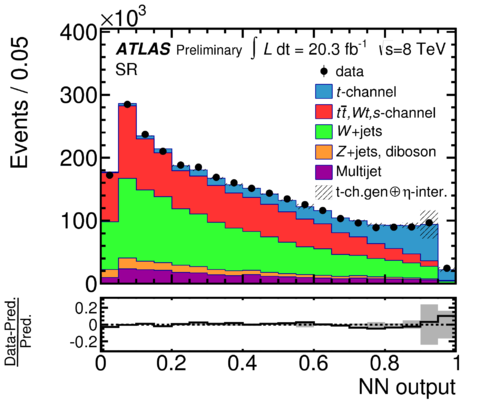}
\includegraphics[width=52mm,height=45mm]{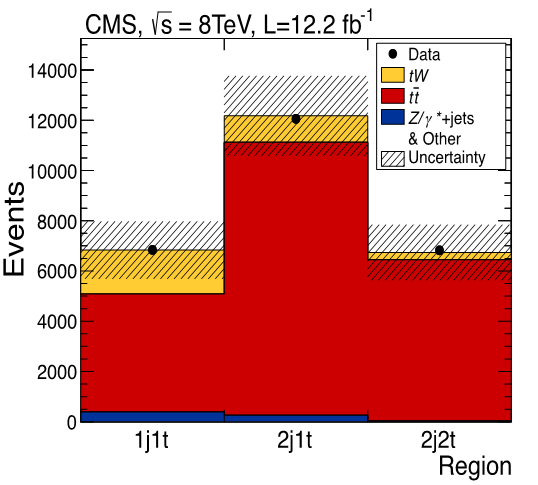}
\includegraphics[width=52mm,height=45mm]{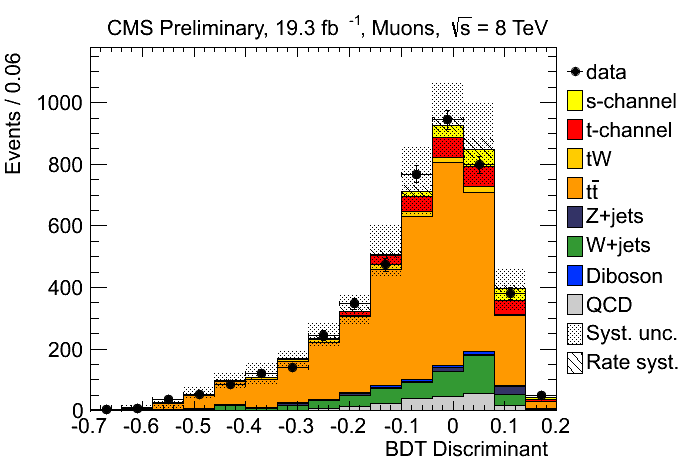}
\end{center}
\caption[]{
Results of $s$-channel, $tW$ production and $t$-channel production for 
(a) $t$-channel at ATLAS, 
(b) $tW$ production at CMS, 
(c) $s$-channel at CMS 
}
\label{fig:lhc}
\end{figure}

ATLAS have searched the single top production in $s$-channel and set upper limit of $\sigma_s <$ 26.5 pb at 95\% CL 
in agreement with the SM of $\sigma_s <$ 20.5 pb.
CMS have measured the $s$-channel production cross section to be $\sigma_s = 6.2^{+8.0}_{-5.1}$ with a significance of 0.7$\sigma$
(0.9$\sigma$ expected) and set upper limit of $\sigma_s <$ 11.5 pb at 95\% CL.

\section{Conclusion}
The most latest single top results at the Tevatron and LHC ere present. 
All single top production processes were observed from the Tevatron and LHC experiments 
by observing the $s$-channel single top production by CDF + D0 experiments, and the $tW$ associated production by CMS experiment
as shown in Table~\ref{tab:summary}.
The all measurements are agreed well with the SM prediction.
More precise measurements for top quark properties will be performed with the single top events at LHC.

\begin{table}[htbp]
\setstretch{1.20}
\centering
\caption[]{Summary of single top quark production cross section at the Tevatron and the LHC}
\label{tab:summary}
\vspace{0.4cm}
\begin{tabular}{|c|c|c|c|c|} \hline

\mco{2}{|c|}{$\sigma$ [pb]}           & $t$-channel                        & $tW$ production   & $s$-channel \\ \hline
\multirow{ 2}{*}{Tevatron (1.96 TeV)} & CDF   & $1.49~^{+0.47}_{-0.42}$ pb & -                 & $1.36^{+0.37}_{-0.32}$ pb \\ \cline{2-5}
                                      & D0    & $3.07~^{+0.53}_{-0.49}$ pb & -                 & $1.10^{+0.33}_{-0.31}$ pb \\ \hline
\multirow{ 2}{*}{LHC (8 TeV)}         & CMS   & $83.6 \pm 7.7$ pb          & $23.4 \pm 5.4$ pb & $<$ 11.5 pb \\ \cline{2-5}
                                      & ATLAS & $82.6 \pm 11.9$ pb         & $27.2 \pm 6.1$ pb & $<$ 26.5 pb \\ \hline
\end{tabular}
\end{table}

\section*{Acknowledgments}

I acknowledge support from the EU community Marie Curie Fellowship Contract No. 302103.
I also thank the organizers of 26th Rencontres de Blois in Blois for an interesting and successful conference.


\section*{References}

\begin{thebibliography}{99}
\bibitem{CDF_observe}T. Aaltonen et al. (CDF Collaboration), {\em Observation of Electroweak Single Top Quark Production}, \Journal{\PRL}{103}{092002}{2009}.

\bibitem{D0_observe}V. M. Abazov et al. (D0 Collaboration), {\em Observation of Single Top-Quark Production}, \Journal{\PRL}{103}{092001}{2009}.

\bibitem{cdfd0} T. Aaltonen et al. (CDF and D0 Collaborations), {\em Observation of $s$-Channel Production of Single Top Quarks at the Tevatron}, \Journal{\PRL}{112}{231803}{2014}.

\bibitem{tevatron_sigma_s}N. Kidonakis, {\em Next-to-next-to-leading logarithm resummation for $s$-channel single top quark production}, \Journal{\PRD}{81}{054028}{2010}.

\bibitem{tevatron_sigma_t}N. Kidonakis, {\em Next-to-next-to-leading-order collinear and soft gluon corrections for $t$-channel single top quark production}, \Journal{\PRD}{83}{091503}{2011}.

\bibitem{cdf_7.5fb} T. Aaltonen et al. (CDF Collaboration), {\em Measurement of the Single Top Quark Production Cross Section and $|V_{tb}|$ in Events with One Charged Lepton, Large Missing Transverse Energy, and Jets at CDF} , arXiv:1407.4031.

\bibitem{D0_s}V. M. Abazov et al. (D0 Collaboration), {\em Evidence for $s$-channel single top quark production in View the MathML source collisions at View the MathML source}, \Journal{\PLB}{726}{656}{2013}.

\bibitem{cdf_s_lvbb} T. Aaltonen et al. (CDF and D0 Collaborations), {\em Evidence for $s$-Channel Single-Top-Quark Production in Events with One Charged Lepton and Two Jets at CDF}, \Journal{\PRL}{112}{231804}{2014}.

\bibitem{cdf_s_vbb}  T. Aaltonen et al. (CDF and D0 Collaborations), {\em Search for $s$-Channel Single-Top-Quark Production in Events with Missing Energy Plus Jets in $p\bar p$ Collisions at $\sqrt s$ =1.96 ??TeV}, \Journal{\PRL}{112}{231805}{2014}.

\bibitem{lhc_sigma}N. Kidonakis, {\em NNLL threshold resummation for top-pair and single-top production}, \Journal{\em Phys. Part. Nucl}{45}{714}{2014}.

\bibitem{cms_t}V. Khachatryan et al. (CMS Collaboration), {\em Measurement of the t-channel single-top-quark production cross section and of the $|V_{tb}|$ CKM matrix element in $pp$ collisions at $\sqrt s$ = 8 TeV}, \Journal{\em J. High Energy Phys.}{06}{090}{2014}.
 
\bibitem{cms_tw}S. Chatrchyan et al. (CMS Collaboration), {\em Observation of the Associated Production of a Single Top Quark and a $W$ Boson in $pp$ Collisions at $\sqrt s$ = 8 TeV}, \Journal{\PRL}{112}{231802}{2014}.
 






\end{thebibliography}
\end{document}